\documentclass[aps,prc,preprint,showpacs,superscriptaddress]{revtex4}  
\usepackage{graphicx}
\usepackage{amsmath}
\begin{document}
\newcommand {\nc} {\newcommand}
\nc {\beq} {\begin{eqnarray}} 
\nc {\eol} {\nonumber \\} 
\nc {\eeq} {\end{eqnarray}} 
\nc {\eeqn} [1] {\label{#1} \end{eqnarray}}   
\nc {\eoln} [1] {\label{#1} \\} 
\nc {\ve} [1] {\mbox{\boldmath $#1$}}
\nc {\rref} [1] {(\ref{#1})} 
\nc {\Eq} [1] {Eq.~(\ref{#1})} 
\nc {\la} {\mbox{$\langle$}}
\nc {\ra} {\mbox{$\rangle$}}
\nc {\dd} {\mbox{${\rm d}$}}
\nc {\cM} {\mathcal{M}} 
\nc {\cY} {\mathcal{Y}} 
\nc {\dem} {\mbox{$\frac{1}{2}$}}
\nc {\ut} {\mbox{$\frac{1}{3}$}} 
\nc {\qt} {\mbox{$\frac{4}{3}$}} 
\nc {\Li} {\mbox{$^6\mathrm{Li}$}}
\nc {\M} {\mbox{$\mathcal{M}$}}
\title{$^{3}{\rm He}(\alpha,
\gamma)^{7}{\rm Be}$ and $^{3}{\rm H}(\alpha,\gamma)^{7}{\rm Li}$
reaction rates and the implication for Big Bang nucleosynthesis in the potential model}
\author {S.A. Turakulov}
\email{turakulov@inp.uz} \affiliation {Institute of Nuclear Physics,
Academy of Sciences, 100214, Ulugbek, Tashkent, Uzbekistan}
\author {E. M. Tursunov}
\email{tursune@inp.uz} \affiliation {Institute of Nuclear Physics,
Academy of Sciences, 100214, Ulugbek, Tashkent, Uzbekistan}
%
\begin{abstract}
The reaction rates of the direct astrophysical capture processes
$^{3}{\rm He}(\alpha, \gamma)^{7}{\rm Be}$ and $^{3}{\rm
H}(\alpha,\gamma)^{7}{\rm Li}$, as well as the abundance of the
$^{7}{\rm Li}$ element are estimated in the framework of a
two-body potential model. The estimated $^{7}{\rm Li/H}$ abundance
ratio of $^{7}{\rm Li/H}=(5.07\pm 0.14 )\times 10^{-10}$ is in a
very good agreement with the recent measurement $^{7}{\rm
Li/H}=(5.0\pm 0.3) \times 10^{-10}$ of the LUNA collaboration.
\end{abstract}
\maketitle
\section{Introduction}

The radiative capture  reactions $^{3}{\rm He}(\alpha, \gamma)^{7}{\rm Be}$ and
$^{3}{\rm H}(\alpha, \gamma)^{7}{\rm Li}$ play an
important role in stellar and primordial nucleosynthesis
\cite{adelber11,fields11}. A quantity of formation of  the $^{7}{\rm
Li}$ isotope after Big Bang  (the cosmological lithiumprimordial abundance) is evaluated from the reaction rates of above
processes.  In last years the problem was discussed extensively from the both experimental and theoretical viewpoints
\cite{coc17,cyburt16}. As a rule, experimental measurements of these reactions in
low-energy region are inaccessible due to strong Coulomb repulsion.
However, very recently in Ref.\cite{takacs15} the $^{3}{\rm
He}(\alpha, \gamma)^{7}{\rm Be}$ astrophysical S-factor was
determined by the indirect method at the Gamow peak energy region.
Astrophysical S-factor $S_{34}$(23$^{+6}_{-5}$ keV)=0.548$\pm$0.054
keV b was obtained by analyzing observed neutrino fluxes from the
Sun within the standard solar model.

On the other hand, recent direct  measurements of the LUNA collaboration \cite{LUNA6Li} and theoretical development  \cite{tur15,tur16,baye18,tur18} within the potential model have done an important step toward the correct estimation of the  primordial abundance of the  $^{6}{\rm Li}$ element.

The aim of present paper is to estimate the $^{3}{\rm He}(\alpha,
\gamma)^{7}{\rm Be}$ and $^{3}{\rm H}(\alpha, \gamma)^{7}{\rm Li}$
reaction rates and to apply the obtained results for the evaluation of  lithium primordial abundance in the framework
of a two-body potential model \cite{tur181} . The first $^{3}{\rm He}(\alpha,
\gamma)^{7}{\rm Be}$ direct capture  process is mostly important for the primordial abundance of the  $^{7}{\rm
Li}$ element and the second  $^{3}{\rm H}(\alpha, \gamma)^{7}{\rm Li}$  process yields a small additional
contribution.

\section{Theoretical model}

\par The reaction rate $N_{A}(\sigma v)$ is estimated with the help
\cite{NACRE99,Fowler}
\begin{eqnarray}
N_{A}(\sigma v)=N_{A}
\frac{(8/\pi)^{1/2}}{\mu^{1/2}(k_{\text{B}}T)^{3/2}}
\int^{\infty}_{0} \sigma(E) E \exp(-E/k_{\text{B}}T) d E,
\end{eqnarray}
where $k_{\text{B}}$ is the Boltzmann coefficient, $T$ is the
temperature, $N_{A}=6.0221\times10^{23}\, \text{mol}^{-1}$ is the
Avogadro number. The reduced mass $\mu=A m_N$ with a reduced mass
number $A=A_1 A_2/(A_1 + A_2)$ for the $^3$He$+\alpha$ and
$^3$H$+\alpha$ systems with $A_1=3$ and $A_2=4$, consequently a
value of $A=12/7$ is fixed. When a variable $k_{\text{B}}T$ is
expressed in units of MeV it is convenient to use a variable $T_9$
for the temperature in units of $10^9$ K according to the equation
$k_{\text{B}}T=T_{9}/11.605$ MeV. In our calculations $T_9$ varies
in the interval $0.001\leq T_{9} \leq 1$.

After substitution of these variables the above integral for the
reaction rates can be expressed as:
\begin{eqnarray}
\label{rate}
 N_{A}(\sigma v)=3.7313 \times 10^{10}A^{-1/2}\,\, T_{9}^{-3/2}
\int^{ \infty}_{0} \sigma(E) E \exp(-11.605E/T_{9}) d E.
\end{eqnarray}

\section{Estimation of  reaction rates for the  $^{3}{\rm He}(\alpha,
\gamma)^{7}{\rm Be}$ process}

\par In Table \ref{ta1} we give theoretical estimations for the $^{3}{\rm
He}(\alpha, \gamma)^{7}{\rm Be}$ reaction rates in the temperature
interval $10^{6}$ K $\leq T \leq 10^{9}$ K ($ 0.001\leq T_{9} \leq 1
$) calculated within two  potential models $V_{M1}^{a}$
and $V_{D}^{a}$. Parameters of these models have been given in Ref.\cite{tur181}.
They yield a good description of the astrophysical S-factor and cross section for the both
$^{3}{\rm He}(\alpha, \gamma)^{7}{\rm Be}$  and $^{3}{\rm
H}(\alpha, \gamma)^{7}{\rm Li}$ direct capture reactions.

\begin{table}[htb]
\caption{Theoretical estimations of the direct $^{3}{\rm He}(\alpha,
\gamma)^{7}{\rm Be}$ capture reaction rate in the temperature
interval $10^{6}$ K $\leq T \leq 10^{9}$ K ($ 0.001\leq T_{9} \leq 1
$)} {\begin{tabular}{@{}cccccc@{}} \toprule $T_{9}$ & $V_{M1}^{a}$ &
$V_{D}^{a}$ & $T_{9}$& $V_{M1}^{a}$ & $V_{D}^{a}$\\ 
\colrule
0.001 & $9.545\times10^{-48}$ & $9.358\times10^{-48}$ & 0.070 & $9.741\times10^{-7}$ & $9.648\times10^{-7}$\\
0.002 & $1.947\times10^{-36}$ & $1.909\times10^{-36}$ & 0.080 & $3.445\times10^{-6}$ & $3.416\times10^{-6}$\\
0.003 & $5.891\times10^{-31}$ & $5.778\times10^{-31}$ & 0.090 & $9.991\times10^{-6}$ & $9.916\times10^{-6}$\\
0.004 & $1.675\times10^{-27}$ & $1.643\times10^{-27}$ & 0.100 & $2.493\times10^{-5}$ & $2.477\times10^{-5}$\\
0.005 & $4.771\times10^{-25}$ & $4.682\times10^{-25}$ & 0.110 & $5.535\times10^{-5}$ & $5.504\times10^{-5}$\\
0.006 & $3.543\times10^{-23}$ & $3.478\times10^{-23}$ & 0.120 & $1.119\times10^{-4}$ & $1.114\times10^{-4}$\\
0.007 & $1.102\times10^{-21}$ & $1.082\times10^{-21}$ & 0.130 & $2.098\times10^{-4}$ & $2.090\times10^{-4}$\\
0.008 & $1.876\times10^{-20}$ & $1.842\times10^{-20}$ & 0.140 & $3.692\times10^{-4}$ & $3.680\times10^{-4}$\\
0.009 & $2.055\times10^{-19}$ & $2.019\times10^{-19}$ & 0.150 & $6.162\times10^{-4}$ & $6.149\times10^{-4}$\\
0.010 & $1.613\times10^{-18}$ & $1.585\times10^{-18}$ & 0.160 & $9.836\times10^{-4}$ & $9.822\times10^{-4}$\\
0.011 & $9.762\times10^{-18}$ & $9.594\times10^{-18}$ & 0.180 & $2.244\times10^{-3}$ & $2.244\times10^{-3}$\\
0.012 & $4.800\times10^{-17}$ & $4.718\times10^{-17}$ & 0.200 & $4.551\times10^{-3}$ & $4.559\times10^{-3}$\\
0.013 & $1.992\times10^{-16}$ & $1.959\times10^{-16}$ & 0.250 & $1.860\times10^{-2}$ & $1.869\times10^{-2}$\\
0.014 & $7.188\times10^{-16}$ & $7.068\times10^{-16}$ & 0.300 & $5.386\times10^{-2}$ & $5.431\times10^{-2}$\\
0.015 & $2.305\times10^{-15}$ & $2.267\times10^{-15}$ & 0.350 & $1.250\times10^{-1}$ & $1.264\times10^{-1}$\\
0.016 & $6.686\times10^{-15}$ & $6.577\times10^{-15}$ & 0.400 & $2.487\times10^{-1}$ & $2.523\times10^{-1}$\\
0.018 & $4.395\times10^{-14}$ & $4.325\times10^{-14}$ & 0.450 & $4.430\times10^{-1}$ & $4.504\times10^{-1}$\\
0.020 & $2.221\times10^{-13}$ & $2.186\times10^{-13}$ & 0.500 & $7.254\times10^{-1}$ & $7.395\times10^{-1}$\\
0.025 & $5.676\times10^{-12}$ & $5.591\times10^{-12}$ & 0.600 & $1.621\times10^{0}$ & $1.660\times10^{0}$\\
0.030 & $6.684\times10^{-11}$ & $6.589\times10^{-11}$ & 0.700 & $3.054\times10^{0}$ & $3.140\times10^{0}$\\
0.040 & $2.405\times10^{-9}$ & $2.374\times10^{-9}$ & 0.800 & $5.115\times10^{0}$ & $5.278\times10^{0}$\\
0.050 & $3.045\times10^{-8}$ & $3.010\times10^{-8}$ & 0.900 & $7.869\times10^{0}$ & $8.145\times10^{0}$\\
0.060 & $2.097\times10^{-7}$ & $2.075\times10^{-7}$ & 1.000 & $1.136\times10^{1}$ & $1.179\times10^{1}$\\
\botrule
\end{tabular} \label{ta1}}
\end{table}

\begin{figure}[htb]
\centerline{\includegraphics[width=10.7cm]{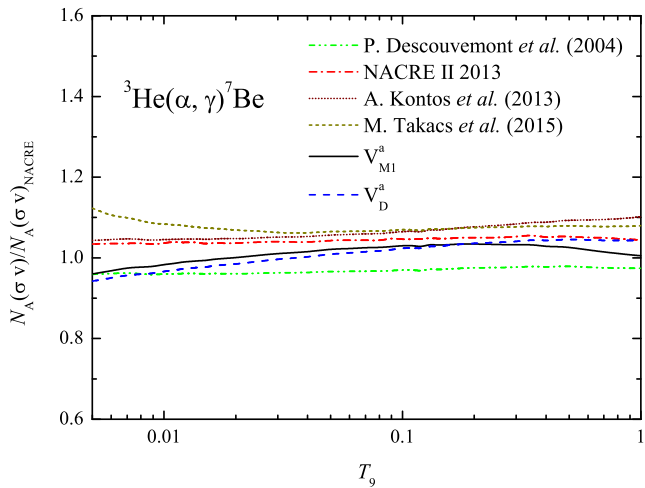}} \vspace*{8pt}
\caption{Reaction rates of the direct $^{3}{\rm He}(\alpha,
\gamma)^{7}{\rm Be}$ capture process within the models $V_{M1}^{a}$
and $V_{D}^{a}$ normalized to the NACRE 1999 experimental data
\label{f3}}
\end{figure}

In Fig. \ref{f3} we display the estimated reaction rates of the
direct $^{3}{\rm He}(\alpha, \gamma)^{7}{\rm Be}$ capture process
within the models $V_{M1}^{a}$ and $V_{D}^{a}$ normalized to the
standard NACRE 1999 experimental data \cite{NACRE99}. For the
comparison we also display the lines corresponding to the results of
Refs.\cite{takacs15,kontos13,desc04} and new NACRE II 2013 data
\cite{NACRE13}. As can be seen from the figure, our results obtained
within the models $V_{M1}^{a}$ and $V_{D}^{a}$  are consistent
with other models.

For the estimation of the primordial abundance of the $^7$Li element
we use  the well known PArthENoPE \cite{Pisanti08} public code,
which however operates only with an analytical form of the reaction
rate dependence on the temperature $T_9$. In this way the theoretical
reaction rate in Table \ref{ta1} is approximated with parameters in
Table \ref{ta2} within 1.02\% (Model $V_{M1}^{a}$) and 0.98\% (Model
$V_{D}^{a}$)  \cite{tur181} using the following analytical formula:

\begin{eqnarray}
\label{analitik}
 N_{A}(\sigma v)=p_0 T_{9}^{-2/3} \exp ( -C_{0} T_{9}^{-1/3}) \times ( 1 + p_1 T_9^{1/3} + p_2
 T_{9}^{2/3}+\\\nonumber
 + p_3 T_{9}+p_4 T_{9}^{4/3}+p_5 T_{9}^{5/3}) + p_6 T_{9}^{-3/2} \exp (-C_{01} T_9^{-1}).
\end{eqnarray}

The coefficients of the analytical polynomial approximation of the
$^{3}{\rm He}(\alpha, \gamma)^{7}{\rm Be}$ reaction rates estimated
with the two different potential models $V_{M1}^{a}$ and $V_{D}^{a}$
are given in Table \ref{ta2} in the temperature interval ($0.001\leq
T_{9} \leq 1 $). For this process $C_{0}=12.813$ and
$C_{01}=15.889$.

\begin{table}[htb]
\caption{Fitted values of the coefficients of analytical approximation
for the direct capture reaction $^{3}{\rm He}(\alpha,
\gamma)^{7}{\rm Be}$} {\begin{tabular}{@{}cccccccc@{}} \toprule
\textrm{Model} &
$p_0$ & $p_1$ & $p_2$ & $p_3$ & $p_4$ & $p_5$ & $p_6$ \\
\colrule

$V_{M1}^{a}$ & $2.691\times10^{6}$ & 8.140 & -26.747 & 43.336 & -35.678 &11.504 & 438.432 \\
$V_{D}^{a}$  & $2.636\times10^{6}$ & 8.128 & -26.559 & 43.243 & -35.608 &11.445 & 453.245 \\

\botrule
\end{tabular} \label{ta2}}
\end{table}

On the basis of the theoretical reaction rates of $^{3}{\rm
He}(\alpha, \gamma)^{7}{\rm Be}$ capture and with the help of the PArthENoPE
\cite{Pisanti08} public code we have estimated the primordial
abundance of the $^7$Li element. If we adopt the Planck 2015 best
fit for the baryon density parameter $\Omega_b
h^2=0.02229^{+0.00029}_{-0.00027}$  \cite{ade16} and the neutron
life time $\tau_n=880.3 \pm 1.1$ s  \cite{olive14}, for the $^7$Li/H
abundance ratio we have an estimate $(4.81 \pm 0.12)\times
10^{-10}$ within model $V_{M1}^{a}$. Model $V_{D}^{a}$ yields an
estimate $(4.92 \pm 0.13)\times 10^{-10}$. As we can see below, these numbers slightly increase
when including a contribution of the $^{3}{\rm H}(\alpha,
\gamma)^{7}{\rm Li}$ direct capture reaction to the $^7$Li/H abundance ratio.

\section{Estimation of  reaction rates for the $^{3}{\rm H}(\alpha,
\gamma)^{7}{\rm Li}$ process}

In Table \ref{ta3} we give theoretical estimations for the $^{3}{\rm
H}(\alpha, \gamma)^{7}{\rm Li}$ reaction rates in the temperature
interval $10^{6}$ K $\leq T \leq 10^{9}$ K ($ 0.001\leq T_{9} \leq 1
$) calculated with the same  potential models $V_{M1}^{a}$
and $V_{D}^{a}$ which were used in the  $^{3}{\rm
He}(\alpha, \gamma)^{7}{\rm Be}$ capture process.

\begin{table}[htb]
\caption{Theoretical estimations of the direct $^{3}{\rm H}(\alpha,
\gamma)^{7}{\rm Li}$ capture reaction rate in the temperature
interval $10^{6}$ K $\leq T \leq 10^{9}$ K ($ 0.001\leq T_{9} \leq 1
$)} {\begin{tabular}{@{}cccccc@{}} \toprule $T_{9}$ & $V_{M1}^{a}$ &
$V_{D}^{a}$ & $T_{9}$& $V_{M1}^{a}$ & $V_{D}^{a}$\\ \colrule
0.001 & $5.594\times10^{-28}$ & $6.129\times10^{-28}$ & 0.070 & $1.326\times10^{-2}$ & $1.461\times10^{-2}$\\
0.002 & $6.284\times10^{-21}$ & $6.886\times10^{-21}$ & 0.080 & $2.838\times10^{-2}$ & $3.129\times10^{-2}$\\
0.003 & $1.612\times10^{-17}$ & $1.767\times10^{-17}$ & 0.090 & $5.382\times10^{-2}$ & $5.936\times10^{-2}$\\
0.004 & $2.251\times10^{-15}$ & $2.468\times10^{-15}$ & 0.100 & $9.313\times10^{-2}$ & $1.028\times10^{-1}$\\
0.005 & $7.496\times10^{-14}$ & $8.218\times10^{-14}$ & 0.110 & $1.501\times10^{-1}$ & $1.657\times10^{-1}$\\
0.006 & $1.081\times10^{-12}$ & $1.185\times10^{-12}$ & 0.120 & $2.285\times10^{-1}$ & $2.524\times10^{-1}$\\
0.007 & $9.073\times10^{-12}$ & $9.949\times10^{-12}$ & 0.130 & $3.322\times10^{-1}$ & $3.671\times10^{-1}$\\
0.008 & $5.233\times10^{-11}$ & $5.739\times10^{-11}$ & 0.140 & $4.648\times10^{-1}$ & $5.139\times10^{-1}$\\
0.009 & $2.296\times10^{-10}$ & $2.519\times10^{-10}$ & 0.150 & $6.300\times10^{-1}$ & $6.968\times10^{-1}$\\
0.010 & $8.193\times10^{-10}$ & $8.987\times10^{-10}$ & 0.160 & $8.310\times10^{-1}$ & $9.195\times10^{-1}$\\
0.011 & $2.487\times10^{-9}$ & $2.729\times10^{-9}$ & 0.180 & $1.352\times10^{0}$ & $1.498\times10^{0}$\\
0.012 & $6.639\times10^{-9}$ & $7.284\times10^{-9}$ & 0.200 & $2.050\times10^{0}$ & $2.272\times10^{0}$\\
0.013 & $1.595\times10^{-8}$ & $1.751\times10^{-8}$ & 0.250 & $4.672\times10^{0}$ & $5.186\times10^{0}$\\
0.014 & $3.515\times10^{-8}$ & $3.857\times10^{-8}$ & 0.300 & $8.661\times10^{0}$ & $9.629\times10^{0}$\\
0.015 & $7.199\times10^{-8}$ & $7.901\times10^{-8}$ & 0.350 & $1.407\times10^{1}$ & $1.566\times10^{1}$\\
0.016 & $1.386\times10^{-7}$ & $1.521\times10^{-7}$ & 0.400 & $2.087\times10^{1}$ & $2.326\times10^{1}$\\
0.018 & $4.407\times10^{-7}$ & $4.838\times10^{-7}$ & 0.450 & $2.899\times10^{1}$ & $3.236\times10^{1}$\\
0.020 & $1.191\times10^{-6}$ & $1.308\times10^{-6}$ & 0.500 & $3.834\times10^{1}$ & $4.284\times10^{1}$\\
0.025 & $8.677\times10^{-6}$ & $9.531\times10^{-6}$ & 0.600 & $6.028\times10^{1}$ & $6.750\times10^{1}$\\
0.030 & $3.919\times10^{-5}$ & $4.307\times10^{-5}$ & 0.700 & $8.585\times10^{1}$ & $9.631\times10^{1}$\\
0.040 & $3.483\times10^{-4}$ & $3.831\times10^{-4}$ & 0.800 & $1.143\times10^{2}$ & $1.284\times10^{2}$\\
0.050 & $1.629\times10^{-3}$ & $1.793\times10^{-3}$ & 0.900 & $1.450\times10^{2}$ & $1.631\times10^{2}$\\
0.060 & $5.242\times10^{-3}$ & $5.773\times10^{-3}$ & 1.000 & $1.773\times10^{2}$ & $1.998\times10^{2}$\\
\botrule
\end{tabular} \label{ta3}}
\end{table}

\begin{figure}[htb]
\centerline{\includegraphics[width=10.7cm]{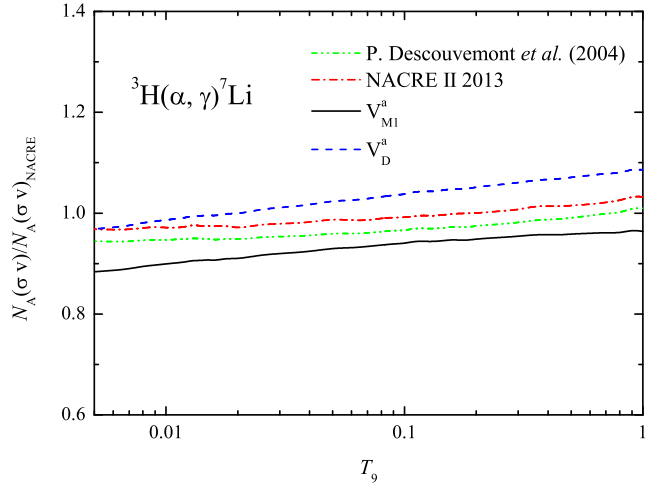}} \vspace*{8pt}
\caption{Reaction rates of the direct $^{3}{\rm H}(\alpha,
\gamma)^{7}{\rm Li}$ capture process within the models $V_{M1}^{a}$
and $V_{D}^{a}$ normalized to the NACRE 1999 experimental data
\label{f4}}
\end{figure}

In Fig. \ref{f4} we display the estimated reaction rates of the
direct $^{3}{\rm H}(\alpha, \gamma)^{7}{\rm Li}$ capture process
within the models $V_{M1}^{a}$ and $V_{D}^{a}$ normalized to the
standard NACRE 1999 experimental data \cite{NACRE99}. For the
comparison we also display the lines corresponding to the results of
Ref.\cite{desc04} and new NACRE II 2013 data \cite{NACRE13}.

The coefficients of the analytical
polynomial approximation of the $^{3}{\rm H}(\alpha, \gamma)^{7}{\rm
Li}$ reaction rates estimated with the potential
models $V_{M1}^{a}$ and $V_{D}^{a}$ are given in Table \ref{ta4} in
the temperature interval ($0.001\leq T_{9} \leq 1 $). For this
process $C_{0}=8.072$ and $C_{01}=3.689$.
In this case the analytical formula (\ref{analitik}) with parameters in
Table \ref{ta4}  reproduces the theoretical reaction rates from the Table
\ref{ta3} within 0.61\% (Model $V_{M1}^{a}$) and
0.62\% (Model $V_{D}^{a}$), respectively.

\begin{table}[htb]
\caption{Fitted values of the coefficients of analytical approximation
for the direct capture reaction $^{3}{\rm H}(\alpha, \gamma)^{7}{\rm
Li}$} {\begin{tabular}{@{}cccccccc@{}} \toprule \textrm{Model} &
$p_0$ & $p_1$ & $p_2$ & $p_3$ & $p_4$ & $p_5$ & $p_6$ \\
\colrule

$V_{M1}^{a}$ & $4.951\times10^{5}$ & 4.034 & -13.145 & 20.845 & -17.355 &5.765 & 39.687 \\
$V_{D}^{a}$  & $5.429\times10^{5}$ & 4.012 & -13.001 & 20.703 & -17.285 &5.747 & 41.245 \\

\botrule
\end{tabular} \label{ta4}}
\end{table}
In Fig. \ref{f4} we display the estimated reaction rates of the
direct $^{3}{\rm H}(\alpha, \gamma)^{7}{\rm Li}$ capture process
within the models $V_{M1}^{a}$ and $V_{D}^{a}$ normalized to the
standard NACRE 1999 experimental data \cite{NACRE99}. For the
comparison we also display the lines corresponding to the
Ref.\cite{desc04} and new NACRE II 2013 data \cite{NACRE13}.

Now  including theoretical reaction rates for the both  $^{3}{\rm He}(\alpha,
\gamma)^{7}{\rm Be}$ and $^{3}{\rm H}(\alpha,\gamma)^{7}{\rm Li}$ capture processes
into the nuclear reaction
network with the help of the PArthENoPE \cite{Pisanti08}public code, we can evaluate
the primordial abundance of the $^7$Li element. Adopting
aforementioned values of the baryon density and the neutron life
time, for the $^7$Li/H abundance ratio we have an estimate $(5.06
\pm 0.13)\times 10^{-10}$ within model $V_{M1}^{a}$. Model
$V_{D}^{a}$ yields an estimation  $(5.08 \pm 0.13)\times 10^{-10}$.
These numbers are slightly larger than the corresponding estimations based on the $^{3}{\rm He}(\alpha,
\gamma)^{7}{\rm Be}$  process exclusively.

\section{Conclusion}

The astrophysical direct capture processes  $^{3}{\rm He}(\alpha,
\gamma)^{7}{\rm Be}$ and $^{3}{\rm H}(\alpha,\gamma)^{7}{\rm Li}$
have been studied in the potential model. The reaction rates and primordial abundance of the
$^7$Li element have been evaluated. It is shown, that the main
contribution to the formation of the $^7$Li  isotope comes from reaction
rates of the $^{3}{\rm He}(\alpha, \gamma)^{7}{\rm Be}$ direct capture process.
Additional contribution is due to the $^{3}{\rm
H}(\alpha,\gamma)^{7}{\rm Li}$ direct capture reaction. For the abundance ratio $^7$Li/H we
have obtained an estimate $(5.07 \pm 0.14)\times 10^{-10}$
consistent with the new data $(5.0\pm 0.3) \times 10^{-10}$ of
the LUNA collaboration \cite{takacs15}.



\begin{thebibliography}{0}    

\bibitem{adelber11} E. G. Adelberger,  {\it et al}., {\it Reviews of Modern Physics\/} {\bf 83}, 195 (2011).
\bibitem{fields11} B. D. Fields, {\it Annual Review of Nuclear and Particle Science\/} {\bf 61}, 47
(2011).
\bibitem{coc17} A. Coc, E. Vangioni, {\it International Journal of Modern Physics E\/} {\bf26}, 1741002
(2017).
\bibitem{cyburt16} R.H. Cyburt, B.D. Fields, K.A. Olive and
T-H. Yeh, {\it Reviews of Modern Physics\/} {\bf 88}, 1 (2016)
\bibitem{takacs15} M.P. Takacs, D. Bemmerer, T. Sz\"{u}cs, and K.
Zuber, {\it Phys. Rev. D\/} {\bf 91}, 123526 (2015).
\bibitem{LUNA6Li} LUNA Collaboration (M. Anders {\it et al}.), {\it Phys. Rev. Lett.\/} {\bf 113}, 042501 (2014).
\bibitem{tur15} E.M. Tursunov, S.A. Turakulov, P. Descouvemont, {\it Phys.
Atom. Nucl.\/} {\bf 78}, 193 (2015).
\bibitem{tur16} E.M. Tursunov, A.S. Kadyrov,  S.A. Turakulov and I. Bray,   {\it Phys. Rev. C\/} {\bf 94}, 015801 (2016).

\bibitem{baye18} D. Baye and E.M. Tursunov. {\it J. Phys. G: Nucl. Part. Phys.\/} {\bf 45}, 085102 (2018)

\bibitem{tur18} E.M. Tursunov, S.A. Turakulov, A.S. Kadyrov and I. Bray.  {\it Phys. Rev. C\/} {\bf 98}, 055803 (2018).

\bibitem{tur181} E.M. Tursunov, S.A. Turakulov and A.S. Kadyrov,  {\it Phys. Rev. C\/} {\bf 97}, 035802 (2018).
\bibitem{NACRE99} NACRE (C. Angulo {\it et al}.),  {\it Nucl. Phys. A\/} {\bf 656}, 3 (1999).
\bibitem{Fowler} W.A. Fowler,G.R.Gaughlan and B.A. Zimmerman, {\it Annu. Rev. Astron.
  Astrophys.\/} {\bf 13}, 69 (1975).
\bibitem{kontos13} A. Kontos, E. Uberseder, R. deBoer, J.
G\"orres, C. Akers, A. Best, M. Couder, M. Wiescher, {\it Phys. Rev.
C\/} {\bf 87}, 065804 (2013).
\bibitem{desc04} P. Descovemont, A.Adahchour, C. Angulo, A. Coc and E. Vangione-Flam, {\it Atomic Data and
Nuclear Data Tables\/} {\bf 88}, 203 (2004)
\bibitem{NACRE13} NACRE II (Y.~Xu {\it et al}.), {\it Nucl. Phys. A\/} {\bf 918}, 61 (2013).
\bibitem{Pisanti08} O.~Pisanti, A.~Cirillo, S.~Esposito, F.~Iocco, G.~Mangano,
G.~Miele, and P. D.~Serpico, {\it Comput. Phys. Commun.} {\bf 178},
956 (2008).
\bibitem{ade16} Planck Collaboration (P. A. R.~Ade {\it et al}.), {\it Astron. Astrophys.\/}
{\bf 594}, A13 (2016).
\bibitem{olive14} Particle Data Group (K. A.~Olive {\it et al}.), {\it Chin. Phys. C\/} {\bf 38}, 090001  (2014).
%

\end{thebibliography}
\end{document}